\documentclass[12pt]{article}
\usepackage{subfigure}
\usepackage{epsfig}
\usepackage{citesort}

\renewcommand{\baselinestretch}{1.87}
\newcommand{\thermotable}{3}
\newcommand{\HBfreqtable}{4}
\newcommand{\figdot}{5}
\newcommand{\figssglobal}{4}
\newcommand{\fighb}{6}

\begin{document}

\begin{titlepage}

\renewcommand{\baselinestretch}{0.8}

\title{
    Solvent and mutation effects \\
    on the nucleation of amyloid $\beta$-protein folding
}

\author{
    Luis Cruz$^{*\P}$,
    Brigita Urbanc$^*$, Jose M. Borreguero$^{*\dagger}$, \\
    Noel D. Lazo$^\ddagger$, David B. Teplow$^\ddagger$ 
    and H. Eugene Stanley$^{*\P}$
    }

\date{}

\maketitle

\begin{center}
{\it {\small 
$*$ Center for Polymer Studies and Department of Physics, Boston
University, Boston, MA 02215.
$\dagger$ Center of Excellence in Bioinformatics, SUNY at Buffalo, NY 14203.\\
$\ddagger$ Department of Neurology, David Geffen School of Medicine at UCLA, Los
Angeles, CA 90095.\\
}}
\end{center}

\begin{flushleft}
\textbf{Classification:} Physical Sciences: Applied Physical Sciences
and Biological Sciences: Biophysics.\\

\textbf{22} text pages (text, references, and legends), \textbf{2} tables, and
\textbf{3} figures.\\
\textbf{250} words in abstract and \textbf{46,722} total number of characters in
the paper.
\end{flushleft}

\begin{flushleft}
$^\P$\textbf{Corresponding Authors: L. Cruz and H.E. Stanley}
\end{flushleft}
Center for Polymer Studies, Department of Physics,\\
590 Commonwealth Ave, Boston University,\\
Boston, MA 02215\\
Phone: 617-353-2617\\
Fax: 617-353-3783\\
Email: ccruz@bu.edu and hes@bu.edu

\begin{flushleft}
{\textbf{Abbreviations:}}
A$\beta$, amyloid $\beta$-protein; 
AD, Alzheimer's disease; 
DMD, discrete molecular dynamics; 
HB, hydrogen bond; 
HCHWA-D, hereditary cerebral hemorrhage with amyloidosis--Dutch type;
MD, molecular dynamics; 
SASA, solvent accessible surface area; 
SB,  salt bridge.
\end{flushleft}

\end{titlepage}

\newpage

\begin{abstract}

Experimental evidence suggests that the folding and aggregation of the amyloid
$\beta$-protein (A$\beta$) into oligomers is a key pathogenetic event in
Alzheimer's disease (AD). Inhibiting the pathologic folding and oligomerization
of A$\beta$ could be effective in the prevention and treatment of AD. Here,
using all-atom molecular dynamics simulations in explicit solvent, we probe the
initial stages of folding of a decapeptide segment of A$\beta$,
A$\beta_{21-30}$, shown experimentally to nucleate the folding process.  In
addition, we examine the folding of a homologous decapeptide containing an amino
acid substitution linked to hereditary cerebral hemorrhage with
amyloidosis--Dutch type, [Gln22]A$\beta_{21-30}$. We find that: (i) when the
decapeptide is in water, hydrophobic interactions and transient salt bridges
between Lys28 and either Glu22 or Asp23 are important in the formation of a loop
in the Val24--Lys28 region of the wild type decapeptide; (ii) in the presence of
salt ions, salt bridges play a more prominent role in the stabilization of the
loop; (iii) in water with a reduced density, the decapeptide forms a helix,
indicating the sensitivity of folding to different aqueous environments; (iv)
the ``Dutch'' peptide in water, in contrast to the wild type peptide, fails to
form a long-lived Val24--Lys28 loop, suggesting that loop stability is a
critical factor in determining whether A$\beta$ folds into pathologic
structures. Our results are relevant to understand the mechanism of A$\beta$
peptide folding in different environments, such as intra- and extracellular
milieus or cell membranes, and how amino acid substitutions linked to familial
forms of amyloidosis cause disease.

\end{abstract}

\begin{flushleft}
\end{flushleft}

\newpage

\section{\noindent{\huge {\bf Introduction}}}

The amyloid cascade hypothesis, first proposed in the early 1990's, posits that
the deposition of amyloid fibrils is the seminal event in the pathogenesis of
Alzheimer's disease (AD)~\cite{CIT:Pik91,CIT:Har91}. However, recent
biophysical, biological, and clinical data indicate that the formation of
oligomeric assemblies much smaller than fibrils may be a key pathologic event
\cite{CIT:Wal97c,CIT:Har97f,CIT:Lam98b,CIT:Kle2001a,CIT:Yon2002a,CIT:Wal2002b,%
CIT:Kir2002a,CIT:Kle2004}. This paradigm shift suggests an attractive
therapeutic approach--prevent or disrupt the assembly of A$\beta$ monomer into
toxic oligomers.  To do so requires an understanding of the molecular dynamics
involved in the transition from non-toxic monomeric to toxic oligomeric states.

Elucidation of the initial events in A$\beta$ monomer folding and assembly is
complicated by the solvent dependence of the process~\cite{CIT:Bar92}. Monomeric
A$\beta$ is largely helical in a membrane or membrane-mimicking environment such
as ionic detergents \cite{CIT:Sti95,CIT:Col98,CIT:Sha99b,CIT:Cre2002}. In
contrast, A$\beta$ monomers in aqueous solution show negligible $\alpha$-helix
or $\beta$-sheet content~\cite{CIT:Bar92,CIT:Zha2000a,CIT:Gur2000}. Studies of
A$\beta_{10-35}$-amide at micromolar concentrations in water show a pH-dependent
folding transition in which the conformation is not helical but contains several
turns and at least two short strands~\cite{CIT:Lee95}. Solution-state NMR
combined with diffusion-ordered spectroscopy indicate that variation of the
anionic strength in the buffer shifts the equilibrium between monomeric
and oligomeric states, possibly allowing for the stabilization of intermediate
structures~\cite{NR05}. Molecular dynamics (MD) studies of A$\beta_{16-22}$
monomer in aqueous urea solution show that increasing the concentration of urea
promotes a transition from a compact random coil to $\beta$-strand structures
\cite{Klimov04}. 

A$\beta$ folding and oligomerization are also influenced significantly by amino
acid structure at specific sites. The dipeptide Ile41-Ala42 at the C-terminus of
A$\beta_{1-42}$ is responsible for the different biophysical behaviors of
A$\beta_{1-42}$ and A$\beta_{1-40}$~\cite{CIT:Bit2003,CIT:Bit2003b}.  Oxidation
of Met35 disrupts A$\beta_{1-42}$ oligomer formation but does not affect
A$\beta_{1-40}$ oligomer formation, consistent with the possibility that there
are structural differences between oligomers of the two alloforms involving
Met35 and neighboring residues~\cite{CIT:Bit2003a}. Simulations of the
oligomerization of A$\beta_{1-40}$ and A$\beta_{1-42}$~\cite{Urbanc04,Urbanc04b}
using discrete MD (DMD) with implicit
solvent~\cite{Smith97,CIT:Dok98a,CIT:Dok2000a,CIT:Smi2001a} are consistent with
{\it in vitro} data, yield new structural predictions, and offer a plausible
mechanistic explanation of the Met35 oxidation experiments \cite{Urbanc04b}. MD
studies show that the presence of glycines induce a transition from an
$\alpha$-helix to a $\beta$-strand conformation in A$\beta_{1-40}$ monomers in
aqueous solution~\cite{Xu05}. Simulation studies of the Dutch
[Gln22]A$\beta_{10-35}$ peptide in water indicate that  interactions between the
peptide and the solvent are stronger in the wild type than in the mutant peptide
\cite{CIT:Mas2003e}.

Recent limited proteolysis experiments on A$\beta_{1-40}$ and A$\beta_{1-42}$,
under conditions favoring oligomerization, identify in both peptides a
protease-resistant segment, Ala21--Ala30. The homologous decapeptide
A$\beta_{21-30}$ shows identical protease resistance~\cite{LGCRT05}. Structure
calculations based on distance constraints from proton solution-state NMR of
A$\beta_{21-30}$ reveal a turn structure at Val24--Lys28. Lazo {\em et
al.}~\cite{LGCRT05} postulate that this structure nucleates the intramolecular
folding of the A$\beta$ monomer and that partial unfolding of the Ala21-Ala30
region may be necessary for the subsequent fibrillization of A$\beta$. The
observations on full-length A$\beta$ and on the A$\beta_{21-30}$ decapeptide are
consistent with previous work showing that peptide fragments containing the
folding nuclei of  globular proteins are, by themselves,
structured~\cite{CIT:Fer99c}.  Furthermore, the structures found in the isolated
folding nuclei are similar to those found in the full-length
proteins~\cite{CIT:Fer99c}.  MD simulations of the folding nucleus, therefore,
provide insights into the earliest events in the folding of the full-length
protein. Borreguero {\em et al.} recently used DMD with implicit solvent and a
united-atom protein model to simulate  folding of the putative A$\beta$ folding
nucleus A$\beta_{21-30}$~\cite{Borreguero05}. The united-atom peptide model
considers explicitly all protein atoms except hydrogen. Important findings in
Borreguero {\em et al.}~\cite{Borreguero05} are: (i) the existence of a loop
that is stabilized by hydrophobic interactions in the Val24--Lys28 region; (ii)
a high degree of flexibility in the termini; and (iii) electrostatic
interactions between the charged groups of Glu22, Asp23, and Lys28 that modulate
the stability of the folded structure.

Here we test whether the stability of the Val24--Lys28 loop described by
Borreguero {\em et al.}~\cite{Borreguero05} persists in simulations that
consider an explicit solvent (all atoms are included in the simulation). In
addition, we determine the effects of solvent alterations on the folding
dynamics and investigate the changes in the dynamics caused by amino acid
substitutions. To this end, we present results of long-time, all-atom MD
simulations of A$\beta_{21-30}$ in explicit water. We also study the dynamics of
the monomer containing the Dutch  [Gln22]A$\beta_{21-30}$ mutation. This
mutation leads to hereditary cerebral hemorrhage with amyloidosis--Dutch type
(HCHWA-D), characterized by extensive deposition of A$\beta_{1-40}$ in
arterioles and small cerebral vessels but only limited senile plaque formation.
{\it In vitro} studies indicate that the Dutch A$\beta$ has a greater tendency
to form protofibrils and fibrils than does the wild type A$\beta$
\cite{CIT:Wal97c,CIT:Mir2000b}. To investigate dynamical differences in the
folding due to changes in the composition and density of the solvent, we study
the system with solvated ions in normal density water and in water with a
reduced density. Our results are consistent with previous results
\cite{Borreguero05} and, in addition, reveal a helix conformation of the
decapeptide when the encompassing water has a lower density than normal water.
Simulations with solvated ionic atoms in the water illustrate how a different
solvent composition can influence the role of salt bridges between the charged
amino acids by strengthening the stability of the Val24--Lys28 loop. In
addition, simulations of A$\beta_{21-30}$ containing the Dutch 
[Gln22]A$\beta_{21-30}$ mutation suggest that the experimentally-observed
increased fibril formation of this peptide may be a consequence of the increased
number of aggregation-prone unpacked conformations.

\section{\noindent{\huge {\bf Methods}}}

{\textbf{MD Simulations}}.
We perform long-time MD simulations of A$\beta_{21-30}$ monomer in water at
normal and reduced density, normal density water with dissociated salt ions, and
the Dutch [Gln22]A$\beta_{21-30}$ monomer in normal density water. We
explicitly consider all atoms with potential energies given by the CHARMM-27
force-field~\cite{CHARMM} using the NAMD package~\cite{KSBBGKPSVS99}. We use the
TIP3P~\cite{JCMIK82} model for the water molecules. We use the NVT ensemble and
confine the system in a box with periodic boundary conditions. We carry out the
MD simulations at a constant temperature of $T=283$K, corresponding to the {\it
in vitro} experiments~\cite{LGCRT05}.

A$\beta_{21-30}$ has the primary structure
Ala-Glu-Asp-Val-Gly-Ser-Asn-Lys-Gly-Ala, where Glu22 and Asp23 are negatively
charged and Lys28 is positively charged, while the Dutch
mutation~\cite{LCFPLDBLF90} substitutes Glu22 by Gln22 (neutral). Following
\cite{LGCRT05}, we block A$\beta_{21-30}$ with NH$_3^+$ and CO$_2^-$ at the N-
and C- termini, respectively. We generate five trajectories with the following
initial conditions: 
\begin{itemize}
\item \ [RC], wild type random coil conformation in normal density water;
\item \ [P1] \& [P2], wild type loop conformations from Lazo {\it et al.}
\cite{LGCRT05} in water with reduced density (corresponding to Families I and II
from Ref.~\cite{LGCRT05}, respectively);
\item \ [DU], Dutch peptide in a random coil conformation in normal density
water;
\item \ and [RCS], wild type random coil conformation in salted water.
\end{itemize}
We choose these five trajectories because they give information about the
dynamics of the wild type A$\beta_{21-30}$ in normal water and water with a
reduced density, starting from experimentally postulated conformations ([P1] and
[P2]) and a control [RC]. These are then compared with the Dutch [DU] and with
the wild type peptide in a salted solvent [RCS].

We solvate each monomer by inserting it in the center of a
previously-equilibrated cube of water molecules of side 43\AA. This insertion
deletes all water molecules overlapping or in close proximity ($<$2.4\AA) to any
of the monomer atoms, resulting in a system with about 2542 water molecules (see
{\it Supporting Information} for details). To maintain system neutrality (for
wild type only), we insert a single Na$^+$ ion far from the monomer. To obtain
the ``salted'' system, we insert 25 dissociated molecules of NaCl, resulting in
a system with 2399 water molecules. We carry out all the insertions and
preparations of the systems using the VMD package~\cite{HDS96}. We use the
Particle Mesh Ewald method~\cite{DYP93}  to calculate the electrostatic
interactions and a cut-off distance of 12\AA\ for the direct electrostatic and
van der Waals interactions. We use a timestep of 1 fs for all simulations and
save configurations (monomer and water) every 2 ps.

We first minimize the energy of the system for 20,000 steps by applying the
conjugate gradient algorithm that relaxes all atoms except the C$_\alpha$ atoms.
Next, we release the C$_\alpha$ atoms and minimize again for another 20,000
steps. Then, we gradually heat the system in the NVT ensemble from $0$ to $283$K
by harmonically constraining all C$_\alpha$ atoms for 100 ps. Following this
step, and still constraining the C$_\alpha$ atoms, we then perform another 200
ps in the NPT ensemble at $283$K, followed by a brief 50 ps NPT simulation with
no constraints to generate the starting configuration for the production runs.
Because of the way that we delete overlapping water molecules in the solvation
step above, these NPT equilibration steps allow for water molecules to fill in
holes left from the insertion of the decapeptide, thus shrinking the size of the
system by about 1\AA\ in all directions (see {\it Supporting Information} for
details). For trajectories [P1] and [P2], we skip the NPT equilibration, not
allowing the system to shrink, thus creating a system in which the effective
density of the water is reduced by about 7.5\%. In Table~\thermotable\ ({\it
Supporting Information}) we list all of the thermodynamic quantities for each of
the trajectories. Total simulation times for the production runs are 102.6~ns
for [RC], 65~ns for  [P1], 83.6~ns for [P2], 80.0~ns for [DU], and 145.0~ns for
[RCS]. Although these simulation times are too small to study the totality of
the folding process of the A$\beta_{21-30}$, they are long enough to exhibit
characteristic elements of the dynamics of the folding for this decapeptide.

{\textbf{Structural Determinants}}. One of the most important quantities
characterizing the structure of A$\beta_{21-30}$ is the distance between atoms.
The notation R($x,y$) signifies the distance R between two specified atoms from
amino acid $x$ and amino acid $y$. The quantities we use to characterize the
structure of A$\beta_{21-30}$ as a function of simulation time are: 
(i) the distance between the two C$_\alpha$ atoms of Ala21 and Ala30
(denoted by R(1,10)) and Val24 and Lys28 (denoted by R(4,8)); the distance
between Lys28(N$_\zeta$) and Glu22(C$_\delta$) (denoted by R(2,8)); the distance
between Lys28(N$_\zeta$) and Asp23(C$_\gamma$) (denoted by R(3,8)),
(ii) the radius of gyration $R_g$ of the monomer given by $R_g^2 \equiv \sum_i
m_i (|\overrightarrow{r_i} - \overrightarrow{r_c}|)^2 / \sum_i m_i$, where
$\overrightarrow{r_c}$ is the center of mass,
(iii) the combined solvent accessible surface area (SASA)~\cite{SOS96} of
the Val24 and Lys28 side chains (denoted by $S$), 
(iv) the hydrogen bond contact maps between amino acids,
(v) the secondary structure propensity for each amino acid,
(vi) the normalized scalar product $D$ between the vector joining
C$_\alpha$ and C$_\beta$ of Val24 and the vector joining C$_\alpha$ and
C$_\epsilon$ of Lys28, and
(vii) the number and type of hydrogen bonds (HB) between atoms in the
decapeptide \cite{HDS96}. 
We calculate secondary structure propensity using the program STRIDE \cite{FA95}
available through the VMD package. A HB exists if the distance
between donor and acceptor is $\leq$3\AA~and the angle $\theta$ between the
donor--hydrogen--acceptor atoms is $160^{\circ}\geq\theta\leq200^{\circ}$. If,
in addition, two hydrogen-bonded atoms are of opposite charge, they form a salt
bridge (SB) \cite{Marqu87}. 

The quantities defined above provide a broad qualitative and quantitative
description of the structure of A$\beta_{21-30}$. In particular, the distance
R(4,8) monitors possible formation of a loop
involving the Val24--Lys28 region and close proximity of the hydrophobic parts
of the Val24 and Lys28 side chains. The secondary structure propensity per amino
acid determines general structure. A decrease of the combined SASA of two
side chains indicates that the side chains are packed, reducing their contact
with the solvent. A reduced normalized scalar product $D$ indicates that the
side chains of Val24 and Lys28 are constrained in their movements, implying that
they  either are bound to the backbone or to other side chains. A small radius
of gyration $R_g$ along with a reduced R(1,10) indicate folded conformations.
Small values of R(2,8) or R(3,8) indicate the possible formation of salt bridges
between Glu22 and Lys28 or Asp23 and Lys28, respectively. Finally, the contact
maps along with the number and type of hydrogen bonds determine the nature and
strength of interactions between different amino acids.

\section{\noindent{\huge {\bf Results}}}

{\textbf{Distances.}} In Fig.~\ref{fig_dists} we show results for R(2,8),
R(3,8), R(4,8), and  $R_g$ as a function of time. As expected, these quantities
fluctuate, but they also show ``events'' that last extended periods of time in
which their values are substantially reduced and their fluctuations are
uncharacteristically small. R(1,10) did not exhibit ``events'' in any of the
trajectories, indicating that the termini are flexible (results not shown). The
graph of $S$ as a function of time has a  close functional resemblance to R(4,8)
(results not shown). This functional resemblance arises because whenever Val24
and Lys28 are in close proximity (R(4,8)$<6.5$\AA), $S$ is reduced from about
575\AA$^2$ (separated and accessible to solvent) to 475\AA$^2$. 

We define ``events'' in Fig.~\ref{fig_dists} by: 
(i) R$^*$(4,8), in which R(4,8) $<$ 6.5\AA\ (Val24--Lys28 loop); 
(ii) R$^*$(2,8) or R$^*$(3,8), in which R(2,8) or R(3,8) $<$ 4.2\AA\  (salt
bridges); 
(iii) S$^*$, in which $S<$ 525\AA$^2$ (Val24 and Lys28 side chain packing);
and 
(iv) R$_g^*$, in which R$_g$ $<$ 6.5\AA\ (peptide packing).
We classify (i) and (iii) as hydrophobic events and (ii) as  electrostatic
events with the formation of  salt bridges (SB).

Events within a trajectory occur at different times and time intervals
(Fig.~\ref{fig_dists}). By adding up the duration in time of these events, we
obtain the total event time per trajectory (Table~\ref{event_table}). Values in
Table~\ref{event_table} are reported in nanoseconds rather than as percentages
because the total simulation times are too short to obtain statistically
significant percentage results. From these data, we see that S$^*$ and
R$^*$(4,8) (hydrophobic events) total times in each trajectory are comparable.
We also see that hydrophobic events last for much longer periods of time than SB
(R$^*$(2,8) and R$^*$(3,8)) events.

We next determine the amount of temporal overlap between pairs of events (Table
\ref{overlap_table}), helpful for establishing correlations between events. We
define the percentage of overlap for event $X$ between two events $X$ and $Y$ as
the time that $X$ and $Y$ overlap divided by the total duration of event $X$,
multiplied by 100. In Table~\ref{overlap_table}, we divide each overlap into two
subcolumns, one for each of the events. In the first overlap column ($S^*$
$\bigcap$ R$^*$(4,8)), we observe that in all trajectories the $S^*$ and
R$^*$(4,8) events are correlated (high \% of overlap), indicating that proximity
of Val24 and Lys28 is linked to an energetically favorable hydrophobic
interaction. In the next column, the values for the overlap between the
R$^*$(4,8) and the SB events (R$^*$(2,8) and R$^*$(3,8) taken together)
demonstrate that, except for trajectories [P2] and [DU], the SB events mostly
occur conditional on an existing R$^*$(4,8) event. Trajectory [P2] does not
allow for this overlap because of its different secondary structure and HB
contacts ({\it Supporting Information}). In trajectory [DU] the lack of a
negative charge in the Gln22 that effectively eliminates all R$^*$(2,8) events
might be key to the lack of R$^*$(4,8) and SB overlaps. Thus, the prominent
overlaps of SB with R$^*$(4,8) suggest that interactions between amino acids in
positions 22, 23, and 28 in the wild type may be intermediary (or necessary) for
extended R$^*$(4,8) events to exist. 

Next, examining R$^*$(4,8) $\bigcap$ R$^*_g$ we observe that for most
trajectories there is a high degree of correlation between the R$^*$(4,8) and 
R$^*_g$ events, indicating that the decapeptide has a smaller effective size
whenever Val24 and Lys28 are in close proximity. Finally, in the last column
(R$^*$(2,8) $\bigcap$ R$^*$(3,8)) we see that the formation of the R$^*$(2,8)
and R$^*$(3,8) salt bridges is mutually exclusive (except in trajectory [RCS]).
This result indicates that while Glu22 (Asp23) interacts with Lys28, Asp23
(Glu22) interacts with other amino acids or water. In contrast, in trajectory
[RCS] the large overlap between the two SB events suggests that the ions in the
solvent facilitate simultaneous formation of both SBs possibly by disrupting
properties of the water molecules surrounding the decapeptide.

{\textbf{Secondary Structure and Packing of Val24 and Lys28.}} In
Fig.~\ref{fig_ss_vs_time} we show the time evolution of the secondary structure
propensity of each amino acid for all the trajectories. As with other structural
determinants, there is a high degree of correlation between these secondary
structures and other events. Typically, R$^*$(4,8) events are correlated with
increased turn propensities that span at least the Gly25--Lys28 region
(Fig.\figssglobal, {\it Supporting Information}). In Fig.~\ref{fig_ss_vs_time}
[P1], the $\pi$-helix correlates with lowered values of all of the R(2,8),
R(3,8), and R(4,8) distances as well as with a very compact conformation with
the lowest R$_g$ (from Fig.~\ref{fig_dists}). We observe a similar behavior in
trajectory [P2] during the helix formation in the later part of the trajectory
(75 to 80 ns). The helices in [P1] and [P2] are both formed under a pre-existing
R$^*$(4,8) event (compare R(4,8) from Fig.~\ref{fig_dists} with
Fig.~\ref{fig_ss_vs_time}). We note that the prominent helices in [P1] and [P2]
both occur in the reduced density water environments whereas in the trajectories
with normal density water there is no observable helix. In trajectory [DU], the
row corresponding to Gly25 (label ``25'' on the y-axis) is less populated (turn)
than the others, suggesting two turn regions separated at Gly25 (see
Fig.~\figssglobal\ [DU], {\it Supporting Information}, for trajectory averages).

We calculate the normalized scalar product $D$ between the vectors formed by the
side chains of Val24 and Lys28. By comparing results on $D$ (Fig. \figdot, {\it
Supporting Information}) with Fig.~\ref{fig_dists}, we observe a persistence of
the relative orientation of the side chains of Val24 and Lys28 during 
R$^*$(4,8) events. These rather stable values for $D$ indicate that during
R$^*$(4,8) events the side chains of Val24 and Lys28 are constrained in their
relative orientation by being bound to atoms in the monomer, increasing packing
and possibly providing stability to the hydrophobic interaction. Trajectory
[DU], however, does not show the same persistence in $D$ values, suggesting that
the lack of R$^*$(4,8) and R$^*$(2,8) events permit an unconstrained motion of
the Val24 and Lys28 side chains.

{\textbf{Hydrogen Bonds.}} An examination of the number of HBs within the
decapeptide as a function of time (Fig.~\fighb\ (upper), {\it Supporting
Information}) reveals that only a handful of short-lived HBs are active during
events. In SB events, there is one HB formed between one of the
Glu22(O$_\epsilon$) (or Asp23(O$_\delta$)) and one of the Lys28(H$_\zeta$). An
exception is trajectory [RCS], in which the two SBs are present at the same
time. The hydrophobic events (R$^*$(4,8) and $S^*$), on the other hand, are not
characterized by unique HBs, but consist instead of different HBs. In
Fig.~\ref{fig_hb_cntmap}, we show contact maps where the strength of contacts is
proportional to the number of HBs between two amino acids. In most trajectories,
Lys28 is the most active amino acid in HB formation. In trajectory [RCS], the
strongest contacts are the salt bridges between Glu22, Asp23, and Lys28.

By analyzing the frequency of formation of individual HBs during hydrophobic
events (Table~\HBfreqtable, {\it Supporting Information}), we determine that the
highest percentage of HBs is formed between atoms belonging to the backbone. The
most prevalent backbone HBs are Lys28(O)-Gly25(H) and Asp23(O)-Lys28(H), and in
trajectory [RC] the longest hydrophobic event that is stabilized by the
Lys28(O)-Gly25(H) also shows the tightest (smallest) loop formation. The
NH$_3^+$ group of Lys28 also participate in SB formation (Table~\HBfreqtable,
{\it Supporting Information}) which is consistent with experimental results
indicating that Lys28 in wild type A$\beta_{21-30}$ is protected from hydrogen
exchange~\cite{LGCRT05}. These results suggest that although transient backbone
HBs may help stabilize an already existing loop, loop formation and long-term
stability might lie in the hydrophobic interactions of Val24 and Lys28 rather
than on specific HBs. On the other hand, loop formation in trajectory [RCS] is
not only driven by hydrophobic interactions but, to a large extent, by SB
interactions, suggested by the large number of SBs in comparison to  backbone
HBs (Table~\HBfreqtable, {\it Supporting Information}).

\section{\noindent{\huge {\bf Discussion}}}

In this work, we present a series of all-atom MD simulations of a decapeptide
folding nucleus of A$\beta$ monomer (A$\beta_{21-30}$) in explicit water and in
water with addition of salt ions. We also present all-atom MD simulations of the
Dutch [Gln22]A$\beta_{21-30}$ peptide. Our results indicate that for all five
trajectories studied (four for wild type and one for Dutch), hydrophobic events,
characterized by packing of the isopropyl group of Val24 and the butyl group of
Lys28, predominate over electrostatic events (SB between Glu22 and Lys28 or
between Asp23 and Lys28).

The hydrophobic events are highly correlated with a smaller value for the radius
of gyration R$_g$, and they last longer than SB events (Tables~\ref{event_table}
and \ref{overlap_table}). In wild type A$\beta_{21-30}$ in normal density water,
results on the distance between termini and lifetimes of backbone HBs indicate
the formation of a semi-rigid loop in the Val24--Lys28 region with highly
flexible termini, in close agreement with DMD studies~\cite{Borreguero05} and
consistent with NMR experiments of A$\beta_{21-30}$~\cite{LGCRT05}. 

In wild type A$\beta_{21-30}$ in reduced density water, the Val24--Lys28 region
also forms a loop that during periods of time may belong to a larger helical
structure (mostly $\pi$- and $\alpha$-helices). The formation of a helix in our
[P1] and [P2] trajectories is not surprising in light of prior studies of
A$\beta$ and other proteins that form helices when the density of water is
effectively reduced by other solvated components
\cite{CIT:Sti95,CIT:Col98,CIT:Sha99b,CIT:Fez2002,CIT:Cre2002}. The formation of
helices in reduced density water is consistent with the possibility that
hydrophobic assembly in proteins is facilitated by removal of water molecules
(``vaporization'') from regions between hydrophobic amino acids
\cite{Chandler02}. Future work in this area will have relevance for
understanding how A$\beta$ folding is affected by different intra- and
extracellular milieus, including endoplasmic and plasma cell membranes,
endosomal/lysosomal compartments, cytoplasm, plasma, and cerebrospinal fluid.

Although in our trajectories with pure normal water the Glu22-Lys28 and
Asp23-Lys28 salt bridges are transient in time, they could be precursors to the
salt bridges proposed in recent molecular dynamics simulations
\cite{CIT:Ma2002b} and modeling based on NMR data \cite{CIT:Pet2002e}, where it
is suggested that the turn or bend in A$\beta$ fibrils is stabilized by a salt
bridge involving Asp23-Lys28. In Lazo {\it et al.}~\cite{LGCRT05}, however, the
R(2,8) and R(3,8) distances (larger than 9\AA) seem to rule out the existence of
the SBs seen here. This finding can be reconciled by considering that NMR
experiments measure averages of these distances and by differences in time
scales, since the simulations here are done in the nanosecond time scale whereas
NMR experiments are performed in the microsecond to millisecond time scales.

In wild type A$\beta_{21-30}$ in normal water with solvated salt ions [RCS], we
observe the longest ``waiting time'' for a hydrophobic event which can be
attributed to the well-known salting-in effect in which a cloud of ions
increases solubility of a protein by lowering its electrostatic free energy
\cite{Creighton93}. In this same trajectory, the coexistence of the Glu22-Lys28
and Asp23-Lys28 SBs during the R$^*$(4,8) event might be a consequence of a
disruption of the HB network of the water by the solvated ions. In this
scenario, ions locally reorient water molecules, changing the HB interactions
between water and Glu22 and Asp23, enhancing the simultaneous formation of these
salt bridges.

In the [DU] trajectory, the lack of hydrophobic and SB events, and the
unconstrained motion of the Val24 and Lys28 side chains, indicate a more
flexible structure than the wild type, in agreement with \cite{CIT:Mas2001g}.
Taken as a whole, results on the [DU] trajectory suggest that the Dutch mutation
alters the folding pathway of the peptide.  According to a postulate of Lazo
{\it et al.}~\cite{LGCRT05}, partial unfolding of the Ala21--Ala30 region of
A$\beta$ should occur prior to fibril formation.  If this is so, then our
observation of the inability of the [DU] trajectory to form a stable Val24-Lys28
loop might explain the higher propensity of the Dutch peptide to form
protofibrils and fibrils.

\section{\noindent{\huge {\bf Conclusions}}}

Our results show that hydrophobic interactions play a crucial role in the
A$\beta_{21-30}$ folding dynamics, assisted by the formation of salt bridges
between the charged amino acids. By performing secondary structure and hydrogen
bond analysis, we find that there is no regular secondary structure or permanent
hydrogen bonding, suggesting that folding involves formation of a  loop
stabilized by the packing of the side chains of Val24 and Lys28. We also show
that by reducing the density of water we may induce formation of a $\pi$-helix.
Interestingly, we find that in normal density water, if the solvent contains
desolvated ions, the salt bridges play a prominent role in the stabilization of
the Val24--Lys28 loop. Finally, we find that for the Dutch
[Gln22]A$\beta_{21-30}$ decapeptide in water, elimination of charge at position
22 disrupts the natural tendency of the monomer to form a long-lived
Val24--Lys28 loop. This substitution likely alters the A$\beta$ folding pathway,
leading to the formation of alternative turn structures, including those
stabilized solely by an Asp23--Lys28 salt bridge.

\bigskip
\noindent {\bf {\large Acknowledgments}}

Financial support was provided by National Institutes of Health grants AG23661,
NS38328, AG18921, and NS44147; and the Bechtel Foundation.

\newpage

% 
% FIGURE CAPTIONS
%

\begin{flushleft}
\textbf{\begin{large}Figure Captions.\end{large}}
\end{flushleft}

\begin{flushleft}
\textbf{Figure 1.} Distances between C$_\alpha$ atoms of Val24 and Lys28
(R(4,8)), charged atoms of Glu22-Lys28 (R(2,8), black), Asp23-Lys28 (R(3,8),
gray), and radius of gyration $R_g$ as a function of  time for each trajectory.
All of the distances are measured in \AA.
\end{flushleft}

\begin{flushleft}
\textbf{Figure 2.} Secondary structure of each amino acid as a function of
 time for each of the trajectories. The amino acid numbering is
sequential from 21 (Ala21) to 30 (Ala30). The colors in the graphs correspond
to: turn (green), bridge (tan), $\alpha$-helix (pink), $\pi$-helix (red), and
coil (white - none of the above).  
\end{flushleft}

\begin{flushleft}
\textbf{Figure 3.} Hydrogen bond contact maps for each of the trajectories. The
level of darkness of each contact is proportional to the number of HBs between
two amino acids (from Table~\HBfreqtable, {\it Supporting Information}). The
grayscale values for each contact map are relative to the maximum and minimum
number of HBs from each trajectory. The strongest contacts per trajectory are:
in [RC] between the Gly25 and Lys28; in [P1] between the Asp23 and Lys28; in
[P2] between the Val24 and Gly29; in [DU] between the Gly25 and Lys28; and in
[RCS] between the Glu22 and Lys28, and Asp23 and Lys28.
\end{flushleft}

\newpage

% ================== TABLES ===========================

%
% TABLE OF ACCUMULATED TIME LENGTHS FOR EVENTS
%
\begin{table}[h]
%\begin{tabular}{|c||c|c|c|c|c|}
%\end{tabular}
\caption{ 
	}
\label{event_table}
\end{table}

\newpage

%
% TABLE FOR EVENT OVERLAP
%
\begin{table}[h]
%\begin{tabular}{|c||c|c||c|c||c|c||c|c|}

%\end{tabular}
\caption{ 
	}
\label{overlap_table}
\end{table}

\newpage

% ================== FIGURES ===========================

%\subsection{Figures}

%
% FIGURE OF DISTANCE PLOTS
%
\begin{figure}[h]
    \centerline{
    }
\caption[]{
	}
\label{fig_dists}
\end{figure}

\newpage

%
% FIGURE OF TIME EVOLUTION OF 2NDARY STRUCTURE
%
\begin{figure}[h]
    \centerline{
    }
\caption[]{
	}
\label{fig_ss_vs_time}
\end{figure}

\newpage

%
% FIGURE OF CONTACT MAPS CONSTRUCTED FROM THE HB CONNECTIVITY
%
\begin{figure}[h]
    \centerline{
    }
\caption[]{
    }
\label{fig_hb_cntmap}
\end{figure}

\end{document}